\documentclass[prl,showpacs,preprintnumbers,twocolumn]{revtex4}

\usepackage{amssymb}

\usepackage{color}

\usepackage{graphicx}

\begin{document}

\title[Intrinsic Tunnel Barriers]{Intrinsic tunneling in phase separated manganites}

\author{G. Singh Bhalla, S. Selcuk, T. Dhakal, A. Biswas, A. F. Hebard}

\email[Corresponding author:~]{afh@phys.ufl.edu}

\affiliation{Department of Physics, University of Florida, Gainesville FL 32611}

\author{}



\begin{abstract}

We present evidence of direct electron tunneling across intrinsic insulating regions in sub-micrometer wide bridges of the phase separated ferromagnet (La,Pr,Ca)MnO$_3$. Upon cooling below the Curie temperature, a predominantly ferromagnetic supercooled state persists where tunneling across the intrinsic tunnel barriers (ITBs) results in metastable, temperature-independent, high-resistance plateaus over a large range of temperatures.  Upon application of a magnetic field, our data reveal that the ITBs are extinguished resulting in sharp, colossal, low-field resistance drops.  Our results compare well to theoretical predictions of magnetic domain walls coinciding with the intrinsic insulating phase. 
\end{abstract}

\maketitle



In ferromagnetic oxides~\cite{ziese143} such as hole-doped manganites, the balancing of electrostatic~\cite{golosov014428} and elastic~\cite{ahn401} energies in addition to competing magnetic interactions may lead to coexisting regions of ferromagnetic metallic \textbf{(}FMM\textbf{)} and insulating phases~\cite{uehara560, zhang805, golosov064404}.  Upon reducing the dimensions of such a system, an increase in the easy{}-axis magnetic anisotropy and a decrease in electrostatic screening~\cite{golosov064404} can create conditions which favor phase separation at the ferromagnetic domain boundaries resulting in novel 
insulating stripe domain walls which allow direct electron tunneling~\cite{golosov064404, rzchowski287, mathur271, milward607}. The manganite (La$_{1-y}$Pr$_{y}$)$_{1-x}$Ca$_{x}$MnO$_{3}$ provides unique opportunities for exploring the formation of such unique domain walls and intrinsic tunnel barriers (ITBs) due to its well-documented micrometer-scale phase separation into FMM and insulating regions~\cite{ahn401, uehara560}. We present here transport properties of thin
(La$_{0.5}$Pr$_{0.5}$)$_{0.67}$Ca$_{0.33}$MnO$_{3}$
(LPCMO) films which, when reduced in dimensions, do indeed exhibit the classic
signatures of tunneling across ITBs separating
adjacent 
FMM regions. 
Further, colossal low field magnetoresistance (MR) measurements suggest that the ITBs coincide with ferromagnetic domain walls, implying that the ferromagnetic domain structure in LPCMO is modified.

To clarify the context and implications of our results, we distinguish
between experiments that are sensitive to the presence of domain walls
coincident with grain boundaries~\cite{gupta15629,hwang2041} or induced at
geometrical constrictions~\cite{mathur6287, wolfman6955, arnal220409}, and our
experiments, where ITBs~\cite{golosov064404,rzchowski287} result from intrinsic phase separation~\cite{ahn401}.
In the former category, enhanced low-field magnetoresistance
observed in polycrystalline ferromagnetic films of
La$_{0.67}$Ca$_{0.33}$MnO$_{3}$ and
La$_{0.67}$Sr$_{0.33}$MnO$_{3}$
is attributed to spin-dependent scattering~\cite{gupta15629}
or tunneling~\cite{hwang2041}
across grain boundaries. Low-field magnetoresistance has also been 
attributed to domain walls artificially induced at microconstrictions
in epitaxial La$_{0.67}$Ca$_{0.33}$MnO$_{3}$
films~\cite{mathur6287} and to domain walls pinned at
nanoconstrictions in patterned epitaxial
La$_{0.67}$Sr$_{0.33}$MnO$_{3}$
films~\cite{wolfman6955, arnal220409}. Below we show that 
choosing a system with the
appropriate material properties and dimensions can give rise to highly
resistive ITB formation in the absence of mechanical
defects.

Calculations incorporating the
double exchange model, which also account for electrostatic interactions
and screening in phase separated ferromagnets show
that narrow stripes of the antiferromagnetic insulating phase (i.e., ITBs)
form at magnetic domain boundaries  
due to enhancement of the
easy-axis anisotropy in ultra-thin (2D) manganite films~\cite{golosov064404}. 
This results in an abrupt change in magnetization between neighboring FMM domains separated by an ITB, in contrast to classical ferromagnets where the direction of magnetization changes over ${\mu}$m length scales near a domain
boundary~\cite{marrows585}. The calculations can
easily be extended to narrow
(1D) geometries such as ours ~\cite{golosov064404}.  We choose LPCMO since near the 
insulator to metal transition temperature, $T_{IM}$, the
FMM phase coexists with insulating phases: the antiferromagnetic
charge{}-ordered insulating~\cite{uehara560} (COI) and the
paramagnetic charge{}-disordered insulating
phases~\cite{podzorov140406, kir024420, lee115118}. Below $T_{IM}$ LPCMO thin films grown on
NdGaO$_3$ are
in a predominantly ferromagnetic state~\cite{dha092404}, making the patterned thin 
films an ideal system for
exploring stripe domain wall and ITB formation. A slight change in the chemistry and doping of the
(La$_{1-y}$Pr$_{y}$)$_{0.67}$Ca$_{0.33}$MnO$_{3}$,
($y$ = 0.5) thin films used in this experiment can drastically alter this phase
composition~\cite{uehara560,ghi184825}. 

\begin{figure}[tbp]
\begin{center}
\includegraphics[angle=0, width=0.35\textheight]{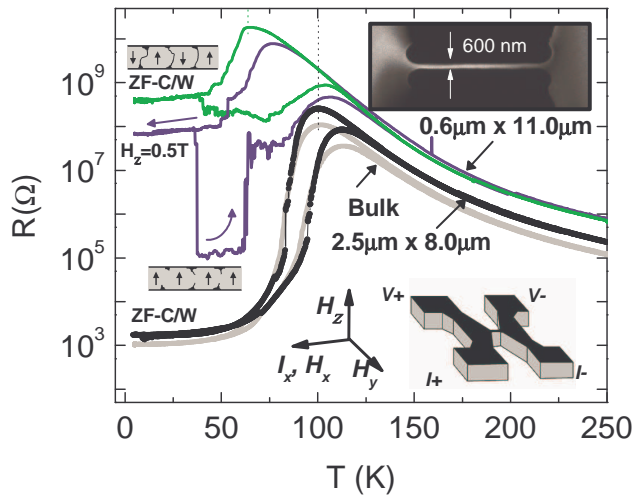}
\end{center}
\caption{\textit{R} vs. \textit{T} upon zero{}-field
cooling and warming (ZF{}-C/W) of bridges patterned from the same LPCMO
film are labeled for the unpatterned film (gray), the 2.5
\textcolor{black}{${\mu}$}\textcolor{black}{m} ${\times}$~8
\textcolor{black}{${\mu}$}\textcolor{black}{m} bridge (black) and the
0.6 \textcolor{black}{${\mu}$}\textcolor{black}{m} ${\times}$~8
\textcolor{black}{${\mu}$}\textcolor{black}{m} bridge (green)
respectively. \ A FC curve for the 0.6
\textcolor{black}{${\mu}$}\textcolor{black}{m} ${\times}$~8
\textcolor{black}{${\mu}$}\textcolor{black}{m} bridge (blue) is also
shown. \ Insulator{}-metal transitions for the two bridges are
indicated by the vertical color{}-coded dashed lines. Lower inset: 
schematic of
the four{}-terminal configuration along with the applied
field directions. 
Upper inset: scanning electron micrograph of
the 0.6~\textcolor{black}{${\mu}$}\textcolor{black}{m} wide bridge.} \label{fig1}
\end{figure}

To fabricate our samples, we epitaxially deposited single crystalline,
30~nm thick
LPCMO films on heated ($820 ^{\circ}$C)
NdGaO$_{3}$ (110) substrates using pulsed laser
deposition~\cite{dha092404}. We first measured (Fig.~1) the
temperature{}-dependent resistance \textit{R}(\textit{T}) of several
unpatterned LPCMO films before standard photolithography and a
manganite wet{}-etch were used to pattern the films into 2.5 x 8
${\mu}$m four{}-terminal bridge structures (see Fig.~1). Measurements
were performed on these bridges before they were reduced
in width using a focused ion beam as shown in the SEM image in
Fig. 1. Care was taken to avoid gallium ion contamination by either
directly depositing a polymer followed by a metal
(70~nm) or just a metal layer on our structure. Four-terminal resistance
measurements were made by sourcing a 1~nA DC current and
measuring the resulting voltages, except for the current-dependent
data shown in Figs.~2 and 3. 

Figure 1 shows \textit{R}(\textit{T}) data for a film patterned into a
bridge geometry of 2.5~${\mu}$m width, which is on the order of
individual domain length scales~\cite{uehara560, zhang805}. In this case,
\textit{R}(\textit{T}) emulates unpatterned thin-film behavior (grey)
with the exception of small step-like features~\cite{zhai167201} below the
insulator-to-metal percolation transition temperature,
$T_{IM}= 100$K~\cite{rai551}. However, when the bridge width is
reduced to 0.6 ~${\mu}$m, transport is clearly dominated by a few
metallic and insulating regions. We attribute the pronounced reduction of
$T_{IM}$ to 64K
for this narrow bridge to
dimensionally{}-limited percolation (from 2D to nearly 1D). Multiple
step-like drops in \textit{R} occur for $T < T_{IM}$
due to
discrete numbers of insulating regions
converting to FMM phase. Recent observations of discrete resistivity
steps on narrow bridges of other mixed phase manganites
also provide evidence of single FMM
and insulating regions spanning the full width of the structure~~\cite{wu214423,zhai167201,yan253121}.

\begin{figure}[bp]
\begin{center}
\includegraphics[angle=0, width=0.40\textheight]{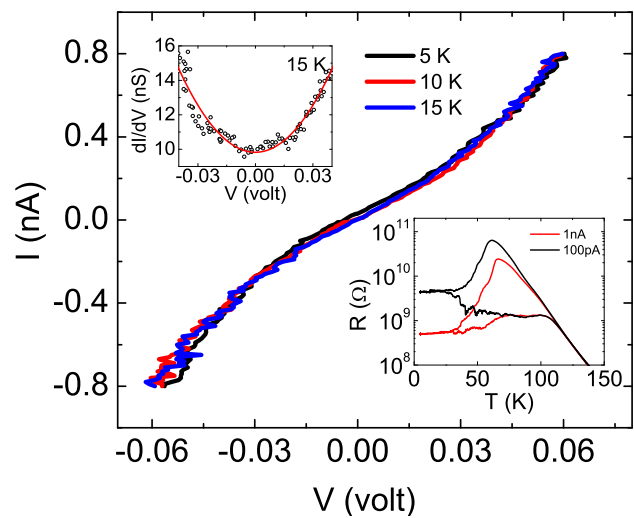}
\end{center}
\caption {$I-V$ characteristics for the
0.6 um wide bridge at the three indicated temperatures all measured
during one cooling cycle. Top inset: d\textit{I}/d\textit{V}{}-\textit{V} curve
at 15 K with a fit (red curve) to the Simmons' model.
Bottom inset: $R(T)$ curves obtained at the indicated currents.} \label{fig2}
\end{figure}

Below \textit{T} ${\approx}$ 50~K, the steps in \textit{R(T)} of the
bridge shown in Fig.~1 cease and a nearly temperature{}-independent
resistance in a supercooled state dominates. Though magnetization
measurements on unpatterned epitaxial LPCMO thin films confirm a
fully ferromagnetic metallic
(3.8${\mu}_B$/Mn) state below
50~K~\cite{dha092404}, it is possible that the narrow geometry of the bridge favors the
formation of an insulating state. However, the temperature range of approximately 50~K
over which this high resistance plateau occurs
cannot be explained by such a scenario since any hopping transport
associated with the insulating COI phase in
LPCMO~\cite{ghi184825, xu2843} would show a pronounced resistance
increase with decreasing temperature.
Additionally, the resistance,
$R \approx 5\times 10^8$~${\Omega}$, of the
zero-field-cooled (ZFC) temperature independent plateau is five
orders of magnitude larger than the quantum of resistance
$h/2e$$^{2}$~=~12.9~k${\Omega}$. By the scaling theory of
localization~\cite{abr673} the large resistance value implies
that for all dimensions, the \textit{T} = 0 state must be an insulator
with infinite resistance, contrary to observation (down to 2~K). We
therefore conclude that transport across the
0.6~${\mu}$m wide bridge is temperature{}-independent \textit{direct}
tunneling through ITBs comprising atomically thin insulating regions.

Also, unique to the 0.6~$\mu$m wide bridge is an abrupt colossal
(thousandfold) drop in resistance near 40~K from the low-temperature
field-cooled (\textit{H}$_{z}$ = 5~kOe) high-resistance plateau
upon field warming (FW) (cf Fig.~1).
Similar but relatively small and smooth drops in $R$ are observed in 
bulk and thin-film LPCMO samples, though the mechanism is unclear~\cite{zhang805}.
To understand this drop in resistance upon FW,
we recall that near $T_{IM}$ in LPCMO
the insulating and metallic regions are not pinned
but evolve in shape and size with changing
temperature~\cite{zhang805,ghi184825}. However, below
$T_B$ $\approx$ 40~K, the
blocking~\cite{ghi184825} or the supercooling glass
transition~\cite{sha224416, wu881} temperature, the phase separated regions are `frozen' in place~\cite{ghi184825,sha224416, wu881}. Thus, upon warming
up again into the dynamic state above
\textit{T}$_B$, the phase separated regions and thus the metastable ITBs are no longer
frozen in space, possibly giving way field-enhanced FMM conversion
of ITBs resulting in a colossal resistance drop. 

To confirm that the temperature{}-independent resistance plateau below
\textit{T}$_{B}$=40~K (see Fig. 1, main text) is due to
ITBs, we measured current-voltage
(\textit{I}{}-\textit{V}) curves at 5~K, 10~K, and 15~K as shown in
Fig. 2. By numerically differentiating the \textit{I}{}-\textit{V}
curve at 15~K, the differential conductance
(d\textit{I}/d\textit{V}{}-\textit{V }) curve shown in the inset of
Fig.~2 is obtained. Assuming one ITB in the bridge, 
the solid red curve was fitted to the data using
the equation, d\textit{I}/d\textit{V} =
\textit{${\alpha}$} + 3\textit{${\gamma}$}\textit{
V}$^{2}$ giving \textit{${\alpha}$} = 9.8(1) ${\times}$
10$^{-9}$ S and \textit{${\gamma}$} ~=
1.0(1)~${\times}$~10$^{-6}$
S/V$^{2}$. Using Simmons' model\cite{simmons1793} and the values for
\textit{${\alpha}$} and \textit{${\gamma}$}, we calculate the average
barrier height $\overline \phi$ to be 0.47~eV and the
barrier thickness (\textit{t}) to be 67~{\AA}. 
Interestingly, our value for $\overline \phi$
is also typical for polycrystalline manganites where
tunneling occurs across a single grain boundary (GB) ~\cite{wes2173, hof681}.
Unlike GBs however, ITBs are
metastable and upon application of a field, bulk
resistivity values are recovered in the bridge, confirming the absence
of GBs in our structure. 

The non{}-linearity of the \textit{I{}-V} curves in Fig. 2 could also
be due to current{}-induced Joule heating. Evidence supporting the
absence of Joule heating is shown in Fig.~2, bottom inset, where \textit{R(T)} is
measured at the indicated currents. Here, $T_{IM}$
increases
with increasing applied current, in contrast to the behavior found by
Sacanell et al. where a decrease in
$T_{IM}$with
increasing applied current is attributed to Joule
heating~\cite{sac113708}.

\begin{figure}[tbp]
\begin{center}
\includegraphics[angle=0, width=0.35\textheight]{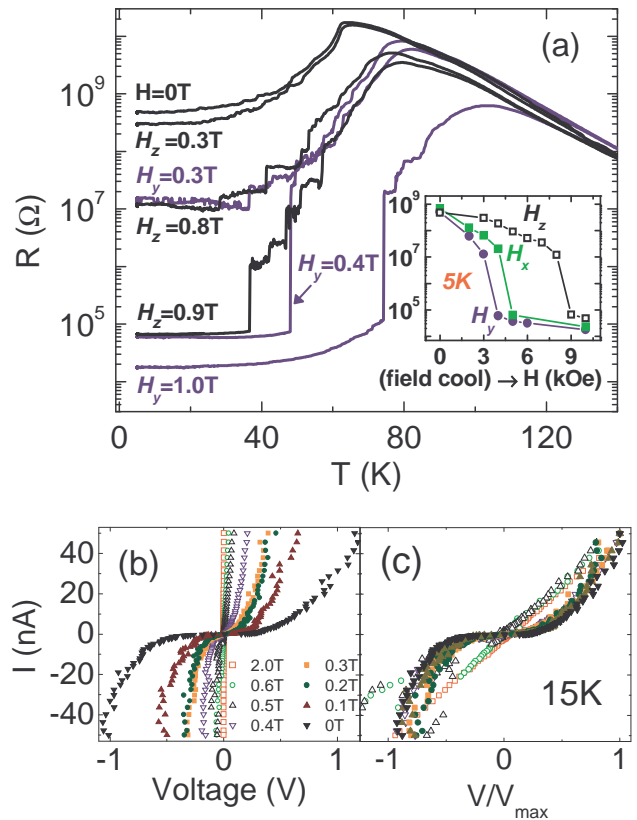}
\end{center}
\caption{a) ZFC and FC resistance transitions are
labeled by the field directions defined with respect to bridge
orientation in Fig.~1. Inset: $R$ measured
at 5~K when the bridge is cooled in separate runs at the indicated
fields. b) \textit{I}{}-\textit{V} characteristics
of the 0.6 ${\mu}$m wide bridge zero{}-field{}-cooled to 15~K, measured
at the indicated values of $H_z$ and (c)
normalized to the voltage, $V_{max}$,
measured at maximum applied current.} \label{fig3}
\end{figure}


Next, we explore the magnetic properties of ITBs.  
If the formation of ITBs is linked to the ferromagnetic domain structure of LPCMO, the magnetic field required for the collapse of the ITB will couple to the intrinsic magnetic anisotropy of the thin film.
Sensitivity of ITBs to magnetic
field direction is verified in Fig. 3a with a subset of the
measured field-cooled $R(T)$ traces for the three field
orientations ($H_x$,$H_y$ and $H_z$) illustrated schematically in
Fig.~1.  The inset highlights the cooling field directional dependence at
5~K. Here, after a resistance of $10^7 \Omega$ is reached,
cooling in a slightly higher field results in an abrupt hundred
fold drop in resistance, suggesting a rapid and sudden
disappearance of the ITB. Clearly, the in-plane fields $H_x$ and $H_y$ are more
effective than $H_z$ in coupling to the magnetization to
reach the low resistance FMM state ($R < 10^5$).  These observations of anisotropic field-induced ITB extinction show a coupling of the ITBs with the magnetic easy axis which lies within the plane of the film for manganites deposited on (110) NdGaO$_3$~\cite{ziese143}. The magnetic
field values required to reach the low resistance state are
on the order of 1~kOe, which though much greater than
the measured coercive field in unpatterned films, are not
atypical for narrow ferromagnetic wires~\cite{yu1859}.  The anisotropic MR associated with the ITBs thus suggests that they may indeed coincide with the FMM domain boundaries forming insulating stripe domain walls.  The necessity of intrinsic phase separation for stripe domain wall formation is apparent when considering recent measurements of notched bridges on non-phase separated La$_{0.67}$Sr$_{0.33}$MnO$_{3}$ which did not show the presence of highly resistive ITBs~\cite{arnal220409}.

Within the context of stripe domain walls, the MR anisotropy arises when spins in neighboring
domains partially align with the field, resulting in reduced
spin{}-dependent scattering at the ITB. The large drop in resistance
signifies a conversion of the ITBs to FMM, analogous to
field induced extinction of Bloch or Neel domain walls.
This notion is confirmed in Fig.~3b
where the magnetic field dependence of the
\textit{I{}-V} curves obtained at 15~K is shown. As $H_z$ increases,
the curvature of the \textit{I{}-V} curves changes
(Fig.~3c). For  fields  sufficiently large enough to drive the bridge into a low resistance state
(${\approx}$9~kOe in Fig. 3a, inset), the \textit{I{}-V} curves become
linear (Fig. 3c) suggesting ITB extinction.

\begin{figure}[tbp]
\begin{center}
\includegraphics[angle=0, width=0.35\textheight]{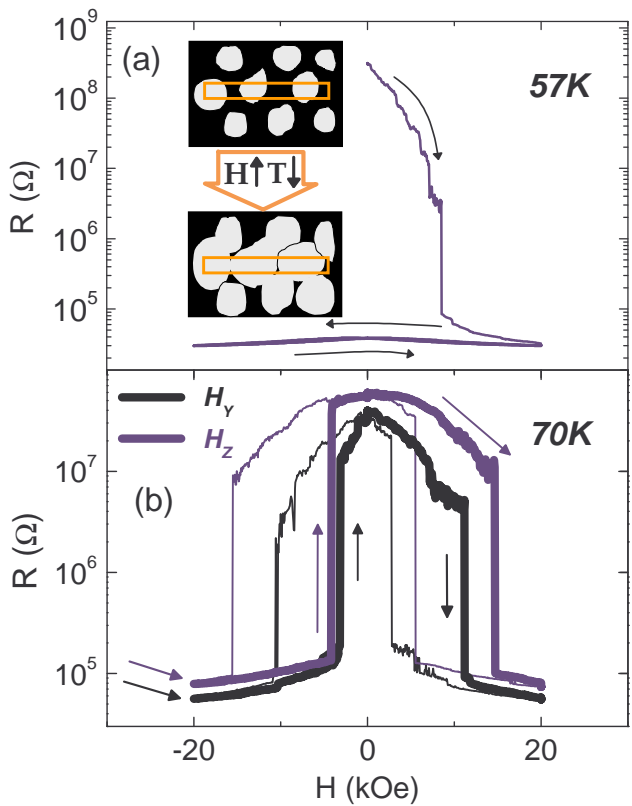}
\end{center}
\caption{Field sweeps ($H$) between ${\pm}$20~kOe
at temperatures a) $T$ = 57~K (below
$T_{IM}$)and b) $T$ = 70~K
(above $T_{IM}$) for both increasing (bold
curve) and decreasing (thin curve) fields:
$H_y$ (black) and
$H_z$ (blue). Inset:
schematic showing the coalescence of FMM (white) regions at the expense of
insulating (black) regions with applied fields or lowering
temperatures. The rectangular overlay depicts the 0.6~$\mu$m bridge.}
\label{fig4}
\end{figure}

The isothermal magnetoresistance curve measured at 57~K in Fig.~4a
follows a similar trend with small incremental decreases in resistance
followed by a large (hundred fold), irreversible, step{}-like drop to a
low resistance state. Each resistance drop can be explained by either
the extinction of an ITB spanning the bridge width, or the incremental
conversion of single ITBs to a FMM state. Qualitatively similar $R$
vs. $H$ curves were measured for all \textit{T} {\textless}
$T_{IM}$ = 64~K.
For temperatures $100 \geq T \geq T_{IM}= 64$K we
observe (Fig. 4b) colossal (hundred fold) field{}-induced resistance
changes which unlike for $T < T_{IM}$ occur at
well-defined anisotropic switching fields (i.e. the resistance can be
switched back to the zero field state upon field removal). 
At the switching field, there is a reduction in the free energy of the
FMM phase and the dominant insulating phase undergoes a first-order
phase transition and a concomitant colossal resistance
drop~\cite{ziese143,mathur271,podzorov140406,ghi184825}. 
Because the changes in the bridge occur more
readily for in{}-plane ($H_y$, blue
curves) easy-axis fields, we speculate that external field{}-induced
spin alignment between neighboring FMM domains enhances the first order
insulator{}-to{}-metal transition, just as ITBs are suppressed below
$T_B$ when neighboring domain spins are
aligned (Fig 2). Below 50~K, the inferred resistance-area
product~\cite{marrows585} (from \textit{R} vs. \textit{H} measurements
and the zero{}-field resistance plateau shown in Fig. 1) of a single
ITB is
${\approx}$10$^{-5}$~$\Omega$m\textsuperscript{2} ~\cite{ziese143} or
more conservatively, for ten ITBs within the bridge,
${\approx}$10$^{-6}$~${\Omega}$m\textsuperscript{2} ~\cite{ziese143}.
These values are 100,000 times greater than
the highest value
(10$^{-11}$~${\Omega}$m\textsuperscript{2})
reported for manganites~\cite{mathur6287, wolfman6955, arnal220409}.

In summary, we have shown that phase-separation in 
manganites is strongly modified in confined geometries 
and leads to the formation of insulating regions thin enough to allow direct electron tunneling, which may also coincide with domain walls separating adjacent FMM domains~\cite{golosov064404}.
Magnetotransport studies of
LPCMO bridge structures with submicron widths less than the average
size of an FMM region enable us to observe tunneling across these metastable ITBs
with record{}-high resistance{}-area products. The high resistance of
the ITBs, is extremely sensitive to temperature and the magnitude and
direction of applied magnetic fields, giving rise to colossal
low{}-field magnetoresistance. In addition to offering rich physical
insights into the formation of ferromagnetic domains in phase separated systems, the presence of
ITBs introduces new opportunities for manipulating
high resistance barriers on the nanometer length scale.

We thank Dmitri Maslov, Pradeep Kumar and Denis Golosov for
discussions. This research was supported by the US National Science Foundation under
grant numbers DMR-0704240 (AFH) and DMR 0804452 (AB).

\end{document}